# MRI correlates of chronic symptoms in mild traumatic brain injury

Cailey I. Kerley[1], Kurt G. Schilling[2], Justin Blaber[1], Beth Miller[3], Allen Newton[2], Adam W. Anderson[2], Bennett A. Landman[1,4,5,] and Tonia S. Rex[3]

[1]Department of Electrical Engineering, Vanderbilt University; [2]Vanderbilt University Institute of Imaging Science, Vanderbilt University; [3]Department of Ophthalmology and Visual Sciences, Vanderbilt University; [4]Department of Biomedical Engineering, Vanderbilt University; [5]Department of Computer Science, Vanderbilt University

## ABSTRACT

Some veterans with a history of mild traumatic brain injury (mTBI) have reported experiencing auditory and visual dysfunction that persist beyond the acute phase of the incident. The etiology behind these symptoms is difficult to characterize, since mTBI is defined by negative imaging findings on current clinical imaging. There are several competing hypotheses that could explain functional deficits; one example is shear injury, which may manifest in diffusion-weighted magnetic resonance (MR) imaging (DWI). Herein, we explore this alternative hypothesis in a pilot study of multi-parametric MR imaging. Briefly, we consider a cohort of 8 mTBI patients relative to 22 control subjects using structural T1-weighted imaging (T1w) and connectivity with DWI. 1,344 metrics were extracted per subject from whole brain regions and connectivity patterns in sensory networks. For each set of imaging-derived metrics, the control subject metrics were embedded in a low-dimensional manifold with principal component analysis, after which mTBI subject metrics were projected into the same space. These manifolds were employed to train support vector machines (SVM) to classify subjects as controls or mTBI. Two of the SVMs trained achieved near-perfect accuracy averaged across four-fold cross-validation. Additionally, we present correlations between manifold dimensions and 22 self-reported mTBI symptoms and find that five principal components from the manifolds (one component from the T1w manifold and four components from the DWI manifold) are significantly correlated with symptoms ($p<0.05$, uncorrected). The novelty of this work is that the DWI and T1w imaging metrics seem to contain information critical for distinguishing between mTBI and control subjects. This work presents an analysis of the pilot phase of data collection of the Quantitative Evaluation of Visual and Auditory Dysfunction and Multi-Sensory Integration in Complex TBI Patients study and defines specific hypotheses to be tested in the full sample.

## 1. INTRODUCTION

Mild TBI (mTBI) is a difficult condition to research; across many studies, even the definition of mTBI injury is disputed [1]. This is unsurprising, however, since across a population of mTBI subjects there is also often a large amount of heterogeneity in both symptoms and imaging findings [2]. Military veterans are particularly susceptible to mTBI due to their frequent proximity to blasts [3]. Many such veterans report experiencing chronic mTBI symptoms, but current clinical magnetic resonance (MR) imaging and computed tomography do not detect any TBI features. The current recommended clinical definition of mTBI, in fact, is a Glasgow Coma Scale score of 13-15 paired with negative imaging findings [1]. Despite this, there are several possible explanations for the chronic dysfunction these veterans are experiencing. Shear injuries, another possible explanation, may manifest on diffusion-weighted MR images (DWI), where they may appear as changes in the geometry, trajectory, and volume of white matter pathways [4]. Based on this alternative hypothesis, in this work we perform a pilot study focused on distinguishing mTBI subjects from healthy controls via multi-parametric MR imaging.

To this end, a set of 1,344 imaging metrics are extracted from DWI and T1-weighted (T1w) MR imaging; these metrics are processed using Principal Component Analysis (PCA) to derive a lower-dimensional representation of the cohort. The

PCA representation of each metric set is employed to train a nonlinear mTBI vs control classifier. Additionally, the connection between imaging metrics and patient symptoms is explored via computing correlations between the PCA representation and self-reported mTBI patient symptoms scores.

## 2. METHODS

### 2.1 Data collection and preprocessing

T1w and DWI scans were acquired for 30 subjects, of whom 22 were controls (no history of TBI nor auditory and visual problems) and 8 were subjects with a history of mTBI (prior mTBI diagnosis confirmed via Electronic Medical Record with a Glasgow Coma Score in the range 13-15). The T1w scans were segmented into 132 brain regions as defined by the BrainCOLOR protocol via multi-atlas labeling as described in [5]. These labels were then registered to the DWI volume space for use in deriving region-based imaging metrics. Both 32 and 64 shell DWI scans were acquired, which were concatenated and corrected for eddy-current distortions and patient movement according to [6].

mTBI subjects filled out a questionnaire including the Neurobehavioral Symptom Inventory, which covers 22 TBI symptoms: dizziness, loss of balance, poor coordination, headaches, nausea, vision problems, light-sensitivity, hearing difficulty, noise sensitivity, numbness, taste or smell changes, appetite changes, poor concentration, forgetfulness, difficulty making decisions, slowed thinking, fatigue, difficulty sleeping, anxiety, feeling depressed, irritability, and frustration [7]. These self-reported symptoms were ranked on a scale of 1 to 5 with regard to its impact on their life since the injury, where 1 was unaffected and 5 was a significant impact on daily life.

### 2.2 Imaging metric extraction

An overview of the derivation of imaging metrics for each MR modality is shown in Figure 1.

#### 2.2.1 DWI metrics

The tractography pipeline was implemented using the MRTrix3 package [8]. The DWI volume was segmented into five tissue-type regions; anatomically-constrained full-brain tractography was then performed [9], and the resulting 10 million streamlines were sifted down to 1 million anatomically-probable streamlines [10]. The DWI-registered BrainCOLOR labels were used to extract four distinct streamline bundles that are associated with the auditory and visual sensory pathways (Figure 2). In the right hemisphere, one bundle connects the thalamus to the superior temporal gyrus, and a second bundle connects the superior temporal gyrus to the calcarine cortex. The third and fourth bundles connect the same structures in the left hemisphere. For each bundle, three metrics were recorded: number of streamlines, average streamline length, and bundle volume, resulting in 12 total connectivity metrics per subject.

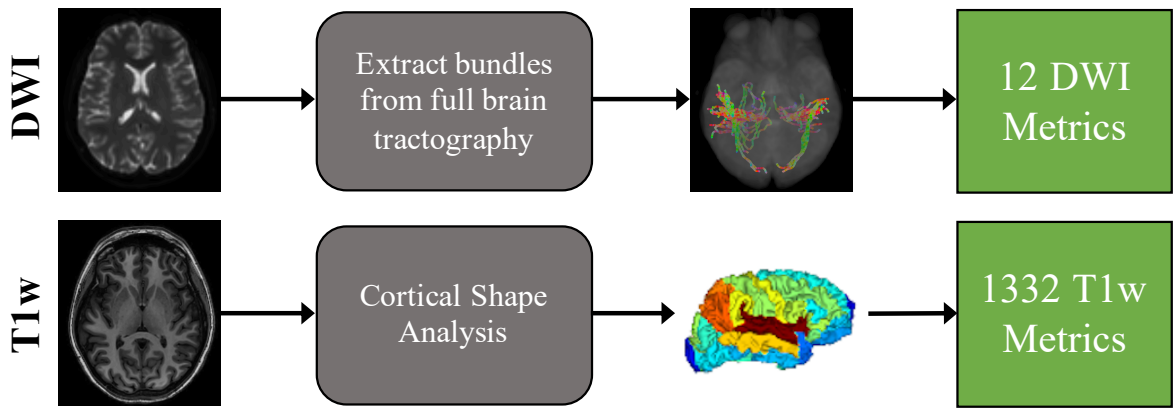

Figure 1. An overview of imaging metric generation is presented. Full-brain tractography is performed on the preprocessed DWI volume, and four streamline bundles are extracted using the BrainCOLOR labels. The number of streamlines, bundle length, and bundle volume are calculated for each bundle, resulting in 12 connectivity metrics per subject. Cortical Shape Analysis is performed on the T1w volume; for each cortical surface region, curvature, shape index, sulcal depth, thickness, shape complexity index, and local gyrification index were calculated both along the region's sulci and averaged across the entire region, yielding 1,332 surface metrics.

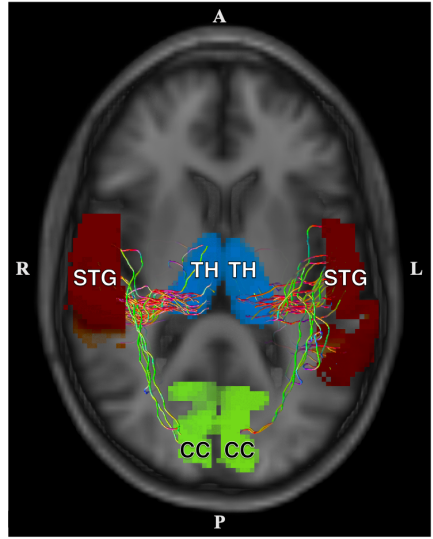

Figure 2. An illustration of the four streamline bundles with the BrainCOLOR regions they connect. In both the right and the left hemispheres, bundles connect the thalamus (TH) to the superior temporal gyrus (STG) and the superior temporal gyrus to the calcarine cortex (CC).

### 2.2.2 T1w metrics

Structural metrics were acquired by first reconstructing the cortical surface and segmenting it into 111 regions via the MaCRUISE pipeline [11]. A cortical shape analysis was then performed on the cortical surface as described in [12]. For each surface region, the mean curvature, shape index, sulcal depth, cortical thickness, shape complexity index, and local gyrification index were calculated both along the sulcal fundic region [12] and averaged across the region as a whole. This resulted in 1,332 structural imaging metrics.

### 2.3 Imaging metric analysis

Figure 3 outlines the analysis of the 1,344 imaging-derived metrics described in section 2.2. The two types of imaging metrics were each analyzed individually to evaluate the effect that each had on the final subject-wise classification. Within each set, the metrics were normalized by calculating the z-score with respect to the mean and standard deviation of the control subjects. PCA was then applied to the z-scores of the healthy controls, producing two individual lower-dimensional PCA spaces (one each for DWI and T1w), which the z-scores of the mTBI subjects were projected into.

Next, each metric set's ability to distinguish between mTBI and control subjects was assessed individually. To this end, an SVM classifier [13] was trained on the PCA space of each metric set combined with subject age. For all SVMs trained, mTBI was defined as the positive class, and control was defined as the negative class. SVMs were trained and validated using four-fold cross validation with a radial basis function kernel. The box constraint and kernel scale hyperparameters were optimized on the training set at each fold using five-fold Bayesian optimization. The PCA spaces of each metric set were iteratively swept, so that a single principal component was added to the SVM training data at each iteration, starting with the first principal component. This sweeping procedure was used to determine how the addition of each component impacted the performance of the SVM classifier.

Once the entire PCA space was swept, the number of components, $C_o$, which produced the most optimal classifier was determined for each metric set. For the purposes of this analysis, "optimal" was defined as the classifier which maximized validation recall (averaged across the four cross-validation folds) based on the fewest number of principal components.

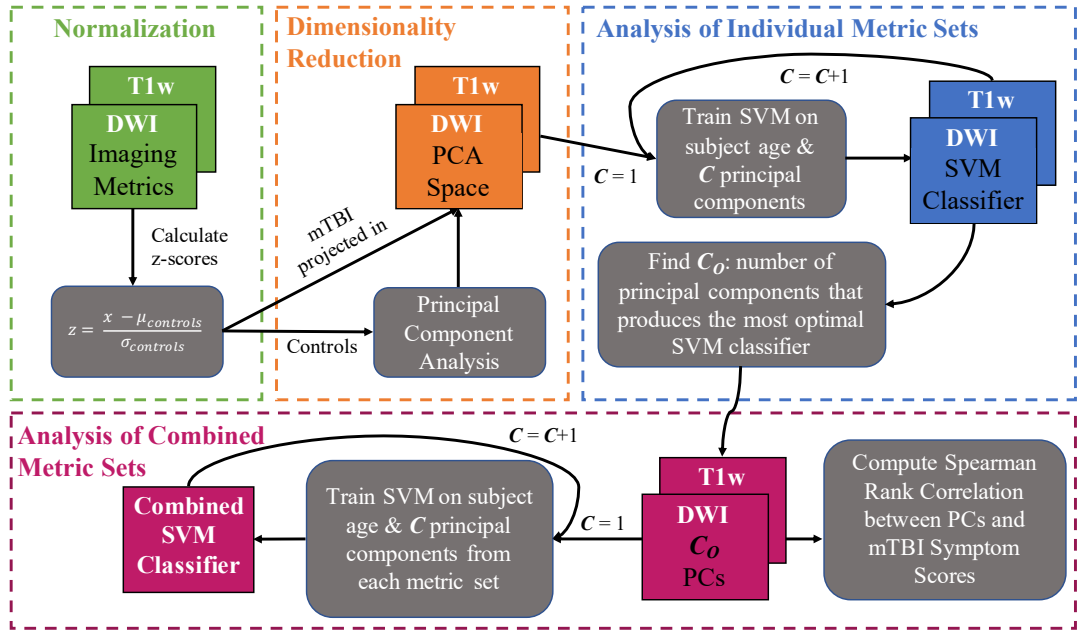

Figure 3. A schematic overview of the imaging metric analysis. First, the imaging metrics are normalized by converting the raw imaging metrics to z-scores using the mean $\mu_{controls}$ and standard deviation $\sigma_{controls}$ of the control subjects. Principal Component Analysis (PCA) is performed using the z-scores of the control subjects, resulting in two lower-dimensional PCA spaces (one for each metric set), which the mTBI subjects' z-scores are projected into. Next, to analyze the metric sets individually, the PCA components of a single set and the subjects' ages are used to train a four-fold cross-validated Support Vector Machine (SVM) to classify subjects as controls or mTBI. Starting with the first principal component, the entire PCA space of each metric set is swept, adding a single component to the SVM at each iteration. After all components have been swept, $C_O$, the number of principal components that produces the most optimal classifier, can be determined for each metric set based on the validation set performance (averaged across the four cross-validation folds). Finally, to analyze the metric sets together, the iterative SVM training process is repeated on the combined set of $C_O$ components from each metric set. In this step, the process starts with the first principal component from each metric set then adds an additional component from each metric set to the classifier at each iteration.

Finally, the $C_O$ components from the DWI and T1w metric sets' PCA spaces were combined, and the iterative SVM training procedure was repeated to analyze how the metric sets might work together to distinguish between mTBI and control subjects. It is important to note that in this combined sweeping, a principal component from both metric sets was added at each iteration, starting with the first principal components from each PCA space. Similar to the individual analysis, the optimal number of components for the combined SVM classifier was determined after the entire set of $C_O$ components from the DWI and T1w metric sets were swept. Additionally, the Spearman rank correlations were calculated between the 22 mTBI symptoms scores and these $C_O$ components from the Combined classifier.

## 3. RESULTS

### 3.1 SVM classifier performance

Figure 4 shows SVM classifier performance as the PCA components are swept for both the individual metric sets and the combined set. This performance is represented by classification accuracy, recall, and specificity, all averaged across the four cross-validation folds. $C_O$ is denoted for each metric set by a vertical bar; for DWI $C_O = 11$, for T1w $C_O = 13$, and for the combined set $C_O = 11$. Note that for the combined set, $C_O$ represents the number of components per metric set (i.e. $C_O = 11$ means that the optimal SVM for the combined set included 11 DWI principal components and 11 T1w principal components). SVM classifier performance at the operating point $C_O$ for each metric set is shown in Table 1.

For both the DWI and T1w metric sets the sweep of the PCA spaces shows that given enough components, the SVM is able to distinguish between mTBI and control subjects, with T1w producing a near perfect classifier. When DWI and T1w were combined, the SVM was again able to distinguish between mTBI and control subjects with performance similar to T1w individually.

### 3.2 Symptom score correlations

Out of the 22 symptom scores, 11 DWI principal components, and 11 T1w principal components, five statistically significant ($p < 0.05$, uncorrected) Spearman rank correlations were found. Table II summarizes these correlations, showing that four of the five significant symptom correlations are related to the DWI metric set, and two of the five symptoms listed are appetite change.

Due to limited sample size and the large number of correlations, these individual tests should be interpreted as exploratory; when false discovery rate correction was applied across all correlations, no tests surpassed the corrected 0.05 threshold.

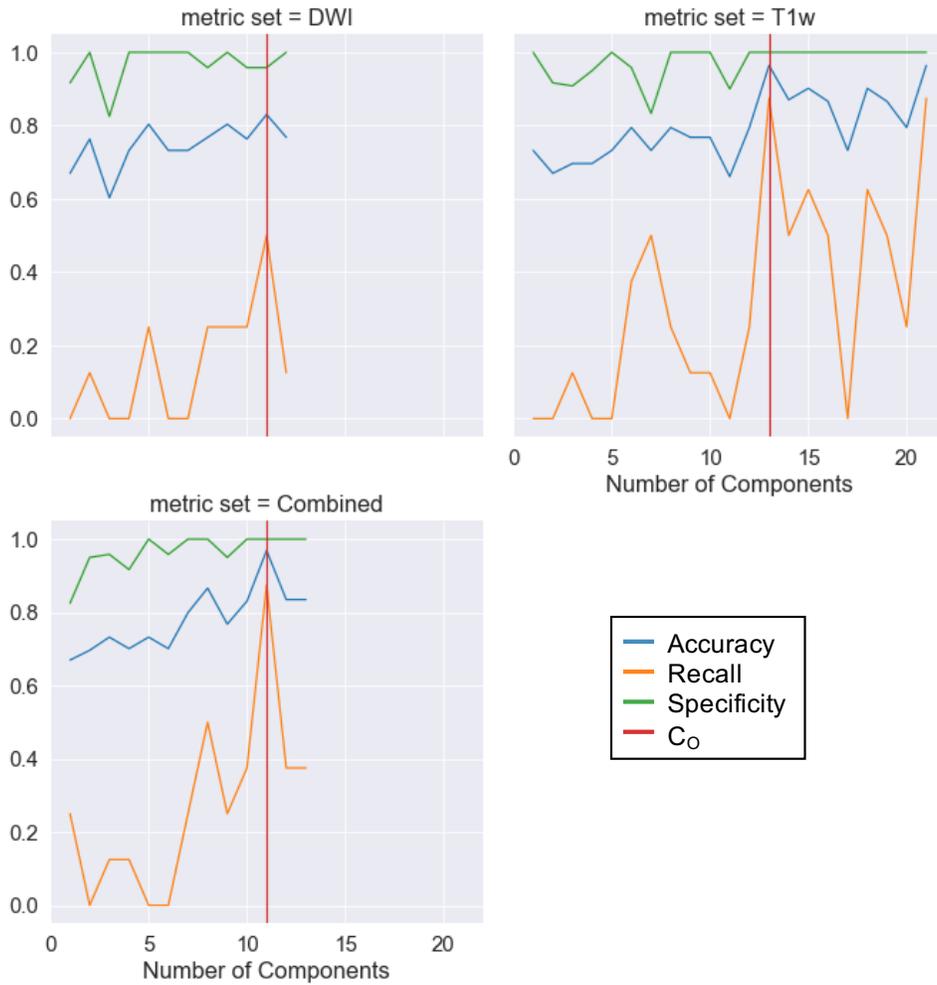

Figure 4. Performance averaged across the 4 folds of individual metric set classifiers and combined metric set classifier as PCA components are added. The optimal operating point is displayed as a red vertical line. The top two plots show that the classifiers trained on the DWI and T1w metric sets individually are able to distinguish between the two classes. The plot in the bottom left shows that the SVM trained on DWI and T1w metric sets combined can also distinguish between the two classes, but not better than the SVM trained only on the T1w metric set.

Table I. Classifier performance for the optimal operating point of each metric set averaged across the four cross-validation folds (values in parentheses are standard deviations)

| Metric Set | C₀ | Accuracy | Recall | Specificity |
|---|---|---|---|---|
| **DWI** | 11 | 0.830 (0.067) | 0.500 (0.354) | 0.958 (0.072) |
| **T1w** | 13 | 0.964 (0.062) | 0.875 (0.217) | 1.000 (0.000) |
| **Combined** | 11 | 0.968 (0.054) | 0.875 (0.217) | 1.000 (0.000) |

Table 2. Significant ($p < 0.05$, uncorrected) correlations found between mTBI symptoms and the PCA components used to train the optimal Combined SVM classifier

| Metric Set | Symptom | Principal Component | Correlation Coefficient | p-value |
|---|---|---|---|---|
| T1w | Appetite Change | 5 | 0.7910 | 0.0196 |
| **DWI** | Appetite Change | 5 | -0.8456 | 0.0178 |
|  | Poor Concentration | 5 | -0.8648 | 0.0357 |
|  | Feeling Depressed | 5 | -0.8225 | 0.0083 |
|  | Frustration | 5 | -0.8225 | 0.0167 |

## 4. DISCUSSION

SVM classifiers trained on the DWI, T1w, and Combined PCA components were all able to distinguish between mTBI and control subjects. The individual T1w and Combined classifiers were both able to achieve near perfect accuracy in this task. It is interesting that the optimal Combined classifier achieved this near-perfect performance using only 11 T1w components, whereas the individual T1w classifier required 13 components to achieve its optimal performance; however, the performance of the two classifiers is too similar to tell whether the T1w or Combined classifier has any true advantage over the other. A second interesting observation is that the DWI metric set produced more significant symptom correlations than the T1w metric set. This suggests that despite its inferior performance in classification, the DWI metric set may still contain some information relevant to mTBI.

## 5. CONCLUSION

The key finding of this work was that the DWI and T1w imaging metrics seem to contain information critical for distinguishing between mTBI and control subjects. For all metric sets, the PCA dimensionality reduction step was performed using data only from controls, yet both the T1w and Combined classifiers achieved near-perfect four-fold cross-validation accuracy. The SVM classification performance indicates that most of the distinguishing information is in the T1w metrics, but the symptom correlations suggest that the DWI metrics may yet prove useful.

In summary, a novel combination of MRI modalities and imaging-derived metrics are presented in an effort to begin characterizing mTBI in MR imaging. Through PCA and SVM, these metrics were leveraged to produce two near-perfect classifiers for a condition that is currently identified by the absence of imaging findings. We conclude, therefore, that the methods described in this work show promise towards characterizing mTBI via MR imaging, but a deeper analysis and larger cohort are needed to clearly determine which individual imaging metrics are contributing the most to subject classification and symptom correlations.

As more data is acquired for this study, we intend to improve the image extraction methods for DWI by including more streamline bundles and extracting more metrics from each bundle (i.e. fractional anisotropy along the bundle, connectivity profile, etc.). Additionally, we plan to deepen the classification and symptom correlation analyses by moving from analyzing whole metric sets to analyzing each metric individually to pinpoint precisely which metrics provide the distinguishing mTBI information.

## ACKNOWLEDGMENTS

Support for this work included funding from DoD grants W81XWH-15-1-0096, W81XWH-17-2-0055, NEI grants R01 EY022349, U24EY029893, P30EY008126, Retired Major General Stephen L. Jones, MD Fund, and Research Prevent Blindness, Inc., as well as NSF CAREER 1452485 (Landman) and NIH grant T32EB001628 (Schilling). This study was in part using the resources of the Advanced Computing Center for Research and Education (ACCRE) at Vanderbilt University, Nashville, TN. This project was supported in part by ViSE/VICTR VR3029 and the National Center for Research Resources, Grant UL1 RR024975-01, and is now at the National Center for Advancing Translational Sciences, Grant 2 UL1 TR000445-06.

## REFERENCES

[1] V. L. Kristman *et al.*, "Methodological issues and research recommendations for prognosis after mild traumatic brain injury: Results of the international collaboration on mild traumatic brain injury prognosis," *Arch. Phys. Med. Rehabil.*, vol. 95, no. 3 SUPPL, 2014.

[2] E. D. Bigler *et al.*, "Heterogeneity of brain lesions in pediatric traumatic brain injury," *Neuropsychology*, vol. 27, no. 4, pp. 438–451, 2013.

[3] C. W. Hoge, D. McGurk, J. L. Thomas, A. L. Cox, C. C. Engel, and C. A. Castro, "Mild Traumatic Brain Injury in U.S. Soldiers Returning from Iraq," *N. Engl. J. Med.*, vol. 358, no. 5, pp. 453–463, Jan. 2008.

[4] G. I. Guberman, J. C. Houde, A. Ptito, I. Gagnon, and M. Descoteaux, "Structural abnormalities in thalamo-prefrontal tracks revealed by high angular resolution diffusion imaging predict working memory scores in concussed children," *Brain Struct. Funct.*, vol. 225, no. 1, pp. 441–459, 2020.

[5] A. J. Asman, A. S. Dagley, and B. A. Landman, "Statistical label fusion with hierarchical performance models.," *Proc. SPIE--the Int. Soc. Opt. Eng.*, vol. 9034, p. 90341E, Mar. 2014.

[6] J. L. R. Andersson and S. N. Sotiropoulos, "An integrated approach to correction for off-resonance effects and subject movement in diffusion MR imaging," *Neuroimage*, vol. 125, pp. 1063–1078, 2016.

[7] K. D. Cicerone and K. Kalmar, "Persistent postconcussion syndrome: The structure of subjective complaints after mild traumatic brain injury," *J. Head Trauma Rehabil.*, vol. 10, no. 3, pp. 1–17, Jun. 1995.

[8] J.-D. Tournier *et al.*, "MRtrix3: A fast, flexible and open software framework for medical image processing and visualisation," *bioRxiv*, p. 551739, 2019.

[9] R. E. Smith, J. D. Tournier, F. Calamante, and A. Connelly, "Anatomically-constrained tractography: Improved diffusion MRI streamlines tractography through effective use of anatomical information," *Neuroimage*, vol. 62, no. 3, pp. 1924–1938, 2012.

[10] R. E. Smith, J.-D. Tournier, F. Calamante, and A. Connelly, "SIFT: Spherical-deconvolution informed filtering of tractograms," *Neuroimage*, vol. 67, pp. 298–312, Feb. 2013.

[11] Y. Huo *et al.*, "Consistent cortical reconstruction and multi-atlas brain segmentation," *Neuroimage*, vol. 138, pp. 197–210, Sep. 2016.

[12] I. Lyu, H. Kang, N. D. Woodward, and B. A. Landman, "Sulcal depth-based cortical shape analysis in normal healthy control and schizophrenia groups," p. 1, 2018.

[13] C. J. C. Burges, "A Tutorial on Support Vector Machines for Pattern Recognition," 1998.